# Design of a novel Electrostatic Ion Storage Ring at KACST

M.O.A. ElGhazaly[1*], S.M.Alshammari[1], C.P.Welsch[2], H.H.Alharbi[1]

[1] *National Center for Mathematics and Physics (*NCMP*),*

*King Abdulaziz City for Sciences and Technology (*KACST*), P.O. Box 6086, Riyadh 11442, Saudi Arabia*

[2] *Cockcroft Institute and the University of Liverpool, UK*

**Abstract.** A new electrostatic storage ring for beams at energies up to 30keV·q is currently under development at the National Centre for Mathematics and Physics (NCMP), King Abdulaziz City for Science and Technology (KACST). The ring design is based on the existing electrostatic storage rings, but stretches significantly beyond them in that it shall form the core of a unique flexible experimental facility at KACST. The lattice of this ring has been designed in a way that enables the use of state-of-the-art experimental methods to study electron-ion, laser-ion, and ion-neutral beams interactions. The lattice design also allows for a future upgrade of the ring to a double storage ring structure that would enable ion-ion beam interactions to be performed. In this paper, we present the design of this ring with a focus on beam dynamics calculations for the 7$^{o}$ single-bend racetrack layout. The study is principally based on the SIMION8 program. We complemented this study further by using purpose-written routine and MAD-X simulation code. An in-depth investigation into beam stability under consideration of non-linear field components in the electrostatic optical elements, is presented. Finally, different working points and stability regions are discussed.

## 1. Introduction

Heavy-ion storage rings have provided a great impetus to research in the field of atomic and molecular collisions. In particular, with the advent of a new generation of storage rings relying exclusively on the use of electrostatic optics, applications toward e.g. biological sciences became possible. A distinctive feature of electrostatic storage rings is that there is in principle no limitation to the mass of the ions that can be stored in such rings [1]. This is because the electric field rigidity scales with the kinetic energy-to-charge ratio *(T/q)* of the particle:

---------------------

* Corresponding author. Tel: +966-1-4814401, Fax: +966-1-4814472. E-mail address: maelghazaly@kacst.edu.sa (M O A El Ghazaly).

$$\rho E = \frac{2T}{q} \qquad (1)$$

As the kinetic energy of the particles (T) depends only from the acceleration voltage ($\Delta U_{acc}$) of the high voltage platform, and the charge state of the particle (q), the electrostatic bending is therefore independent of the ion's mass *m* and this is in contrast to the magnetic field rigidity, which is mass dependent:

$$\rho B = \frac{1}{q}\sqrt{2mT} \qquad (2)$$

Electrostatic rings thus enable the storage of heavy molecular ions and allow for in-depth studies of macromolecules, such as clusters or large bio-molecular systems, including proteins, amino acids and even DNA fragments. This opens up new opportunities in interdisciplinary fields such as e.g. biophysics and biochemistry. In addition to the unique advantage of not being limited in the mass of stored ions, electrostatic rings also benefit from the absence of hysteresis effects and magnetic remanence, ease of operation, since the number of ion optical elements is rather small and compact dimensions, keeping overall costs low. Moreover, such rings can be cooled down to very low temperatures, e.g. by cryo-cooling thereby providing some control of the black-body radiation induced photo-decay of weakly-bound negative ions [2].

The first electrostatic storage ring dedicated to research in atomic and molecular physics appeared in 1998. The ELectrostatic Ion Storage ring in Aarhus (ELISA) is a compact racetrack shaped ring of about 7 m circumference [1]. The concept of ELISA relies on the exclusive use of electrostatic fields to act on the stored ion beam. The second electrostatic ring was built at the high energy accelerator research organization (KEK) in Tsukuba, Japan [3] and was an improved version of ELISA, with cylindrical deflectors to bend ions instead of spherical deflectors as used initially in ELISA. The KEK-ring also features a racetrack lattice of two super periods with mirror symmetry. Each super period consists of one $160^o$ bending deflector, inserted equidistantly between two $10^o$ Parallel-Plate Deflectors (PPD), and two quadrupole doublets for beam focusing. Both rings operate at a similar energy range between 20-25 keV·q. Although similar in design, both rings follow a rather different research program, with groups in Aarhus focusing on measurements of lifetimes of metastable states, cluster dynamics, and photo-excitation of chromophores of photoactive proteins [4-9]. In terms of interaction with electrons, ELISA relies on the use of an electron target in a $90^o$ crossed-beam configuration [10], whereas the KEK-ring has a longitudinal electron target used in merged beam geometry, which can also be used to realize some beam cooling [11]. Because of the rather short interaction length between the electron beam and the stored ion beam, the achievable cooling effect is, however, somewhat limited. The integrated electron target/cooler is a unique feature of the KEK ring and enabled a broad research program in electron scattering on bio-molecular ions [12-13]. The electron target at ELISA has been extensively used with exciting experiments to probe the existence of multiple charge resonance states and their stabilization upon hydration by single water molecules [14-15]. Comprehensive reviews on the research program at both rings can be found in [2, 16-18]. A third ring, also relying on a similar racetrack layout has been developed by the atomic physics and cluster chemistry groups at Tokyo Metropolitan University (TMU) and was constructed by Toshiba [19]. The TMU E-ring, in short, follows a different vacuum and cooling concept, in which the electrodes and shields surrounding the nominal trajectory of the stored ions are kept at liquid nitrogen ($LN_2$) temperature. Thereby, the ions cool down mostly to the vibrational ground state [19-20]. Taking advantage of this distinctive feature, the TMU E-ring has enabled a variety of exciting experiments to be performed, mostly oriented to the structure and dynamics of clusters, as well as lifetime measurement of multiply charged metastable states [21].

Based on the success of the broad experimental program that became accessible through the above-mentioned storage ring facilities, a number of additional electrostatic storage rings have been designed and are presently being built up or commissioned: An electrostatic storage ring with a design energy of up to 50 keV·q is presently being commissioned at the Institut für Kernphysik (IKF) at the University of Frankfurt, Germany [22]. The Frankfurt Low Energy Storage Ring or FLSR in short is optimized to focus the stored beam down to a very small cross section below 1 mm² at the interaction points. In particular, FLSR features a rather large deflection in a PPD with a deflecting angle of 15°. A ground-breaking concept in terms of both flexibility and cooling has come together with the development and construction of a double cryogenic storage ring structure; DESIREE (Double ElectroStatic Ion Ring ExpEriment), at Stockholm University, Sweden [23-24]. The individual storage rings in DESIREE feature a racetrack layout similar to that of ELISA and derived versions. DESIREE is a unique facility that will enable the interactions between negative and positive ions to be studied at low and well-defined internal temperatures and relative energies down to 10 K and 10 meV, respectively. The Cryogenic electrostatic Storage Ring (CSR) that is being constructed at the Max-Planck Institute for Nuclear Physics in Heidelberg, Germany [25-26] stretches significantly beyond the preceding rings in e.g. the availability of state-of-the-art techniques and experimental methods such as a reaction microscope or the use of an ultra-cold electron target, and gas-jet target. CSR features a quadratic shaped layout with four straight sections interconnected by 90° split-bending corners. In particular, it is designed to operate under extreme vacuum conditions and ultra-low temperatures down to $\sim$2K, and so it will be a unique facility in that sense. At the time being, an even miniaturized electrostatic storage ring has been developed and built at the University of Lyon in France [27]. This tabletop Mini-Ring features a rather different shape and follows a different optical concept, making use of conical electrostatic mirrors instead of bending deflectors. More recently, a novel compact cryogenic ion storage ring is now under construction at RIKEN in Japan [28]. This RIKEN mini cryogenic ring is designed to operate at very low temperatures, down to the temperature of liquid helium. Another Ultra-low energy electrostatic Storage Ring (USR) will form a central installation in the future Facility for low-energy Antiproton and Ion Research (FLAIR), at GSI, Germany [29] and a proposal for a new electrostatic storage ring is currently under discussion at NASA's Jet Propulsion Laboratory [30].

At KACST, a highly flexible storage ring for beams at energies up to 30keV·q is presently being developed to complement this spectrum of facilities. With the right experience gained during the work at the pioneering Danish storage ring ELISA [14-15], we have been working, here at KACST, on the development of a facility for atomic and Molecular collisions around an electrostatic storage ring. In this context, a specific design for a highly flexible storage ring relying on the distinctive features of the existing designs such as of ELISA, FLSR, and DESIREE [31] has been carried out. Preliminary lattice studies, using MAD-X code, of the basic conceptual design have been done [32-33]. The conceptual design of such a ring relies on the distinctive features of the existing designs such as of ELISA, FLSR, and DESIREE [31]. Preliminary lattice studies, using MAD-X code, of this basic conceptual design have been done [32-33]. Here, we have used the SIMION program to reproduce the exact shape of electrodes and simulate the fields in the most realistic distribution. Furthermore, changes in the optics configuration have been imposed by the mechanical design and geometrical constraints, and so lattice calculations had to be refined. Thereby, the present design study complements preliminary studies and finalizes the design.

Although the design concept follows the known preceding electrostatic storage rings, our design has a number of distinct features that will make this ring a truly unique experimental facility: It has been designed to become the core of an experimental facility that will allow combining many different, yet complementary beam techniques. The association of such complementary techniques is projected to enable, for the first time in electrostatic storage ring experiments, absolute measurement of cross-sections for atomic and molecular collision processes [34]. Furthermore, our design features a rather significantly small deflection in a PPD with a deflecting angle of $7^{o}$. The following sections present the details of the design of this $7^{o}$ electrostatic storage ring. The following sections present the details of this $7^{o}$ design of the KACST electrostatic storage ring.

The following sections present the details of the design of this $7_{o}$ electrostatic storage ring

## 2. Linear Ion Beam Optics

All the above storage rings rely on the exclusive use of electrostatic fields for bending and focusing the beam. These are generated in parallel plate deflectors, cylinder deflectors and electrostatic quadrupoles, and one therefore refers to *linear beam optics*, i.e. a storage ring where no higher order (sextupole, octupole, etc) fields are used. This does, however, not mean that there are no higher order multi-pole fields in such storage rings, as fringe fields at the entrance and exit region of every optical element may have significant higher order components. Their effect on the stored ion beam needs to be carefully studied as it can dramatically limit beam lifetime and reduce the ring acceptance and dynamic aperture. In the present study, we addressed the question whether fringe fields compromise the ability of an electrostatic ring to store beams of low-energy up to 30 keV$\cdot$q. We first demonstrated that the ring can efficiently store low-energy beams in spite of existing fringing fields.

In order to minimize fringe field effects, grounded shields are normally installed at the entrance and exit of every deflector. This allows also controlling the geometrical length of the electrodes so that the effective path length corresponds to the design values [31]. Thereby one can maximize the acceptance and dynamic aperture of the ring. The only exception to this design concept is the design of the 7° parallel-plate deflectors, which are described in more detail in section 3.1.

Fringe fields in an electrostatic bend cause a change in the kinetic energy of the particles, which in return get focused at different image points and one refers to this effect as chromatic aberrations [35]. Spherical aberrations are rather related to the geometry of the lens and the topology of the lens field. Both, spherical and chromatic, aberrations can cause limitations in dynamic aperture of an electrostatic ring [36]. Still, in the context of comparing electrostatic and magnetic optics, the aberrations generated by electrostatic optics are often larger than those caused by their magnetic counterparts, see e.g. [36]. We therefore explored the aberrations induced by quadrupoles as they are the main focusing lenses in an electrostatic ring. An in-depth investigation into these non-linear effects, complementing commonly used design procedures, had not been realized for any of the existing or planned electrostatic storage rings.

The field components that act on a charged particle in linear beam optics approximation are dipole fields to bend particles along a radius $\rho$ and quadrupole fields to focus the beam with a focusing strength $k$. In this case, the transverse electric field in x direction can be written as:

$$E_x = [E_x]_{x=0} + \left(\frac{dE_x}{dx}\right)_{x=0} x$$

$$= E_{xo}\left(1 - \frac{n}{\rho}x\right) \qquad (3)$$

Here, the index $o$ refers to an on-axis particle, $x$ and $\rho$ are the horizontal displacement from the design trajectory and the local radius of curvature, respectively. From (1) one knows that $E_{xo}=2T/(q{\cdot}\rho)$ and n defines the field index. On this basis, the general equations of motion in the horizontal and vertical planes can be written as [31]:

(Horizontal) $$\frac{d^2x}{ds^2} + \left[\frac{(3-n)}{\rho^2} - k\right]x = \frac{1}{\rho}\frac{\Delta T}{T} \qquad (5)$$

(Vertical) $$\frac{d^2y}{ds^2} + \left[\frac{(n-1)}{\rho^2} + k\right]y = 0 \qquad (6)$$

where $s$ is the position along the beam axis and $\Delta T$ the energy spread.

Since the distribution of the field strengths $1/\rho$ and $k$ change along the nominal orbit, it is impossible to derive direct analytical solutions for the trajectory equations (5) and (6). The layout therefore is started by using an abrupt-field boundaries-model for all ion optical elements where the field begins and ends abruptly at the entrance and exit of each electrode. In this idealization, one can represent the whole ring in matrix form by multiplying the different transfer matrices of all ion optical elements. This approach is integrated in established simulation packages, such as COSY Infinity [37] and MAD-X [38], and can also be easily implemented in purpose-written codes. These studies then need to be complemented by a more detailed analysis including realistic field distributions, using for example programs such as SIMION [39, 40], OPERA3D [41] or CPO [42]. The present design study is based on SIMION version 8, supported by purpose-written routine and the MAD-X code.

## 3. Ion Optical Elements

The optical elements that make up the lattice of the of the ring design consist of dipoles for beam bending and quadrupoles for beam modulation, interconnected by field-free drift regions. In the design of an electrostatic storage ring, bending is either realized by cylinder deflectors or pair of deflecting plates. The latter are also used for beam injection and extraction. Field-free regions are used for decoupling the fields from different optical components, as well as for accommodating beam diagnostics devices and experimental setups. In the following paragraphs, an overview of the design of all elements is given.

### 3.1. Parallel-Plate Deflector

Parallel-Plate Deflectors (PPDs) have been designed for beam injection and small angle deflection; $\Theta<15^o$. Here, the design is optimized for deflecting the beam by an angle of $\Theta=7^o$. This is significantly smaller than the deflecting angle in existing rings, which range between 10° and 15°. This allows us to lower the switching voltage during beam injection and hence reduces switching times and injection effects. A second advantage in lowering the deflection angle is that the effect of aberrations may be reduced [43]. As the deflectors stretch into a curved vacuum chamber it was not possible to design a shield geometry similar to all other optical elements. Even so, the support of the deflecting plates itself provides a limiting effect on the fringe field at the beam entrance side of this deflector. An illustration of such a PPD is shown in Fig. 1, and its main design parameters are summarized in Table 1.

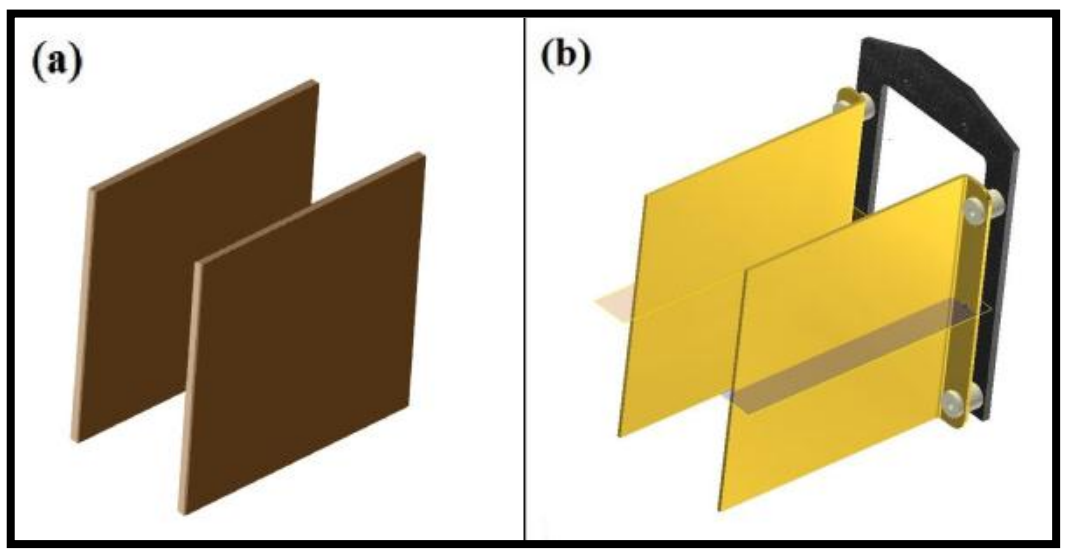

Figure 1: Design of the Parallel-Plate Deflector (PPD). (a) Design used in SIMION simulations (b) PPD with the support of the deflecting plates as designed by Inventor 3 of AutoDesk.

### 3.2. Cylinder Deflector

Using (5) and (6) the equation of motion in a cylinder deflector, where n=1, can be written as:

$n=1$ describes the field in a cylindrical deflector and $n=2$ the field in a spherical deflector.

(Horizontal) $$\frac{d^2x}{ds^2} + \frac{2}{\rho^2}x = \frac{1}{\rho}\frac{\Delta T}{T} \qquad (7)$$

(Vertical) $$\frac{d^2y}{ds^2} = 0 \qquad (8)$$

It can be seen from (7) that there is a weak focusing effect, which contributes to the Betatron oscillations in the horizontal plane, while (8) shows that the cylinder deflector acts as a field-free drift region in the vertical plane. This is in contrast to spherical deflectors where a focusing effect is found in both transverse planes. Measurements showed that in this case non-linear fields lead to a significant reduction in dynamic aperture and even beam losses at higher beam intensities [44]. In this design of a 7$^{o}$ deflection in a PPD, a cylinder deflector with a central radius of $\rho$=*250 mm* bends the beam by an angle of β=166°. Its main nominal parameters are also included in Table 1. An illustration of 166° cylinder deflector set-up, including its shields, is shown in Fig. 2. This figure also shows the vertical steerer located 10 mm upstream of the entrance shield of the cylinder deflector.

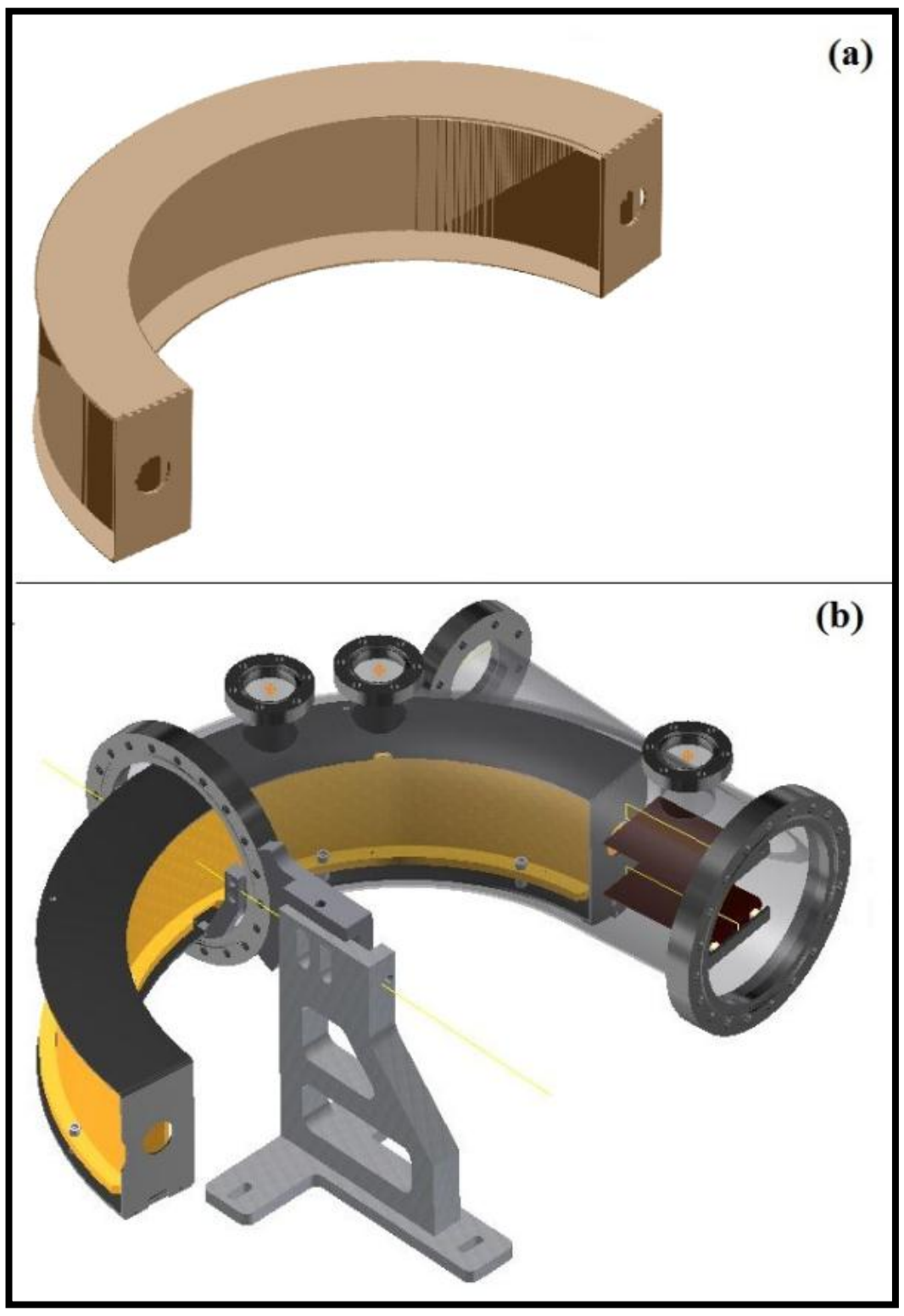

Figure 2.: Layout of the cylinder deflector set-up including longitudinal and lateral grounded shields (a) Design used in SIMION simulations (b) Cylinder deflector set-up inserted in its vacuum chamber together with the Vertical Steerer (the pair of parallel-plates in brown) as designed by Inventor 3.

The position of the effective field boundaries have been calculated to evaluate the effects caused by fringe fields on the beam. Among others, this allows the effective length of the deflector to be determined. The effective deflecting angle of the cylinder deflector can then be estimated from the following parametric function [45]:

$$\Delta\beta = \frac{G_o}{\pi\rho}\left[ln\left[\frac{4}{(x^2+4)}\right] - 2x\,arcos\left[\frac{x}{\sqrt{(x^2+4)}}\right]\right] \qquad (9)$$

Here, $\Delta\beta$ is the extension of the bending angle due to the fringe field, $G_o$ is half of the width of the separation gap between the electrodes, $\rho$ is the radius of curvature of the design orbit, i.e. the central radius of curvature in the centre between both electrodes, and $x=D/Go$ is the separation distance between the electrodes and the grounded shield, normalized to $G_o$.

Function (9) is illustrated in Fig. 3. As can be seen, it yields a bending extension of $\Delta\beta=1.02^o$. In terms of optical structure, this extension of the bending angle needs to be taken into account when determining the geometrical length of the deflector electrodes. Final adjustment will then be realized during operation by small changes in electrode voltages. Moreover, formula (9) allows us optimize the geometry of the entire deflector setup and in particular the best shield configuration that minimizes or even fully compensates the fringe fields effect [43]. One can see from Fig. 3 that the bending extension reduces to $\Delta\beta\sim0.54^o$, if $D$ is reduced to *5 mm*. This then requires a rather small geometrical adjustment that can normally be compensated by slight adjustment of the voltage on the electrodes.

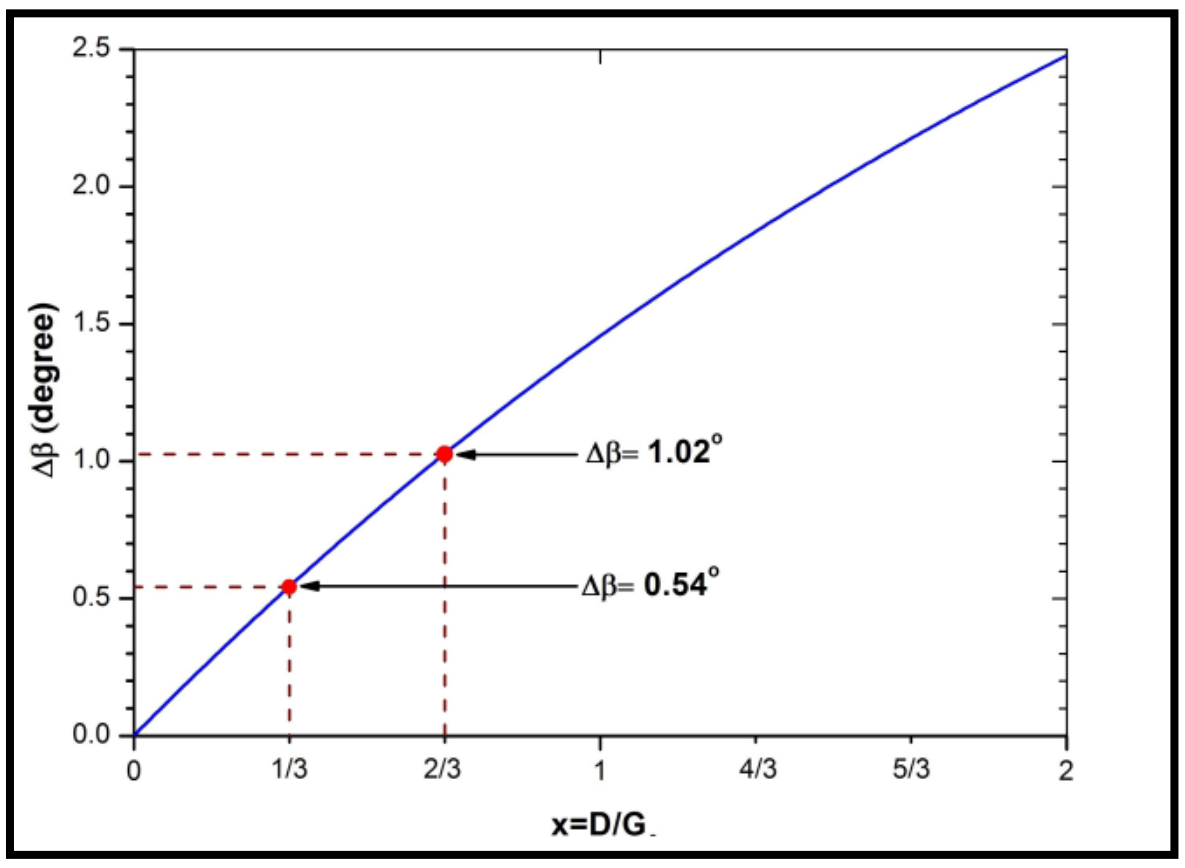

Figure 3: Extension of the bending angle caused by the fringe field as a function of $x=D/G_0$.

### 3.3. Quadrupole Doublets

Similar to magnetic quadrupoles, electrostatic quadrupoles defocus in one plane, while focusing in the other. Consequently, in the design of an electrostatic storage ring, quadrupole doublets have to be used to achieve a net focusing effect in both transverse planes. Design of the quadrupole doublets that will be used in the KACST storage ring is shown in Fig. 4., and their nominal parameters are listed in Table 1.

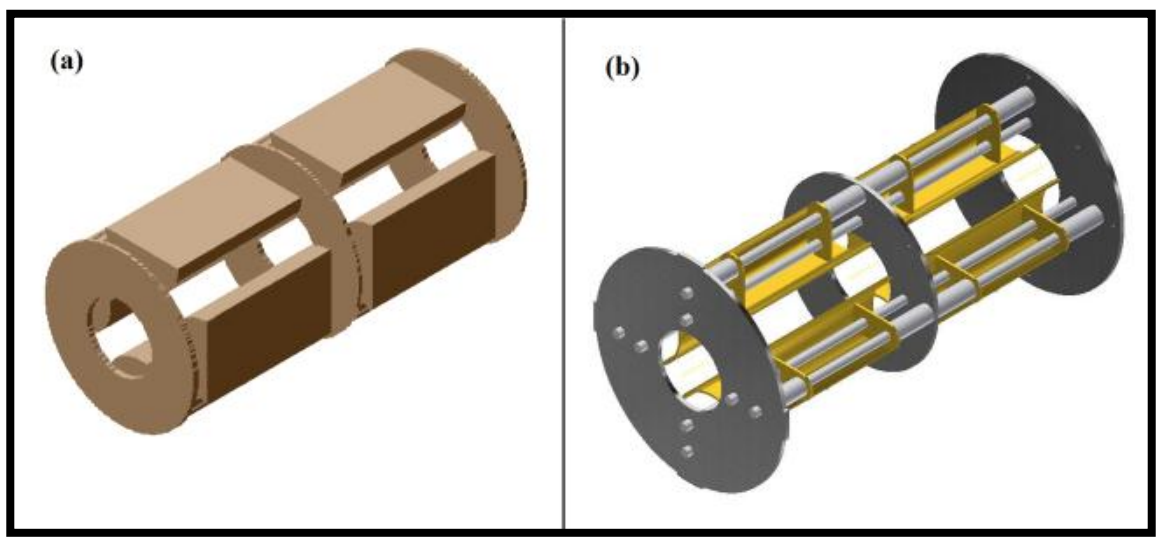

Figure 4: Design of the quadrupole doublet with grounded shields. (a) As included in SIMION simulations. (b) Technical design drawing as designed by Inventor 3 of AutoDesk.

In order to decouple the fields in the doublet, the quadrupoles are designed to be set apart from each other by 33 mm. In addition, the design includes a central shield inserted equidistantly between the quadrupoles. This reduces any possible coupling effects between the two quadrupoles and allow for tuning the focusing strength in the horizontal and vertical planes individually**.** To quantitatively determine the coupling effect between the quadrupoles in a doublet, the field distribution was calculated for the following voltage settings:

($V_{QF}$, $V_{QD}$) = (405, 360), (405, 405), (405, 0) and (0, 360). $V_{QF}$ and $V_{QD}$ are in units of volts as applied to the focusing quadrupole (QF) and defocusing quadrupole (QD), respectively. These settings correspond to a stable working point (405, 360), a configuration where both quadrupoles have identical, but opposite strengths, the individual action of QF, and finally the individual action of QD. Fig. 5 illustrates the simulated field distributions at a position *x=5 mm* from the optical axis of the quadrupole. The illustration shows that the zero-field crossing is exactly in the middle of the central shield, when $V_{QF}$ and $V_{QD}$ are equal in absolute value, while it moves away from the middle of the shield by ~0.5 mm at the working point. This variation is the fact of the difference between $V_{QF}$ and $V_{QD}$ and is thus too small to be seen in the figure. A very small coupling effect between the two quadrupole fields can be observed in a region of about 60 mm around the central shield.

Quadrupoles are responsible for modulating the transverse beam shape in the ring and are the dominant factor in causing betatron oscillations in both, the horizontal and vertical planes. For this reason, the quadrupoles are the most important ion optical element to be considered when it comes to determining the stability of the beam storage in an electrostatic ring. In the following the simulated quadrupole strengths are numerically scanned over a wide range to determine the theoretical stability regions.

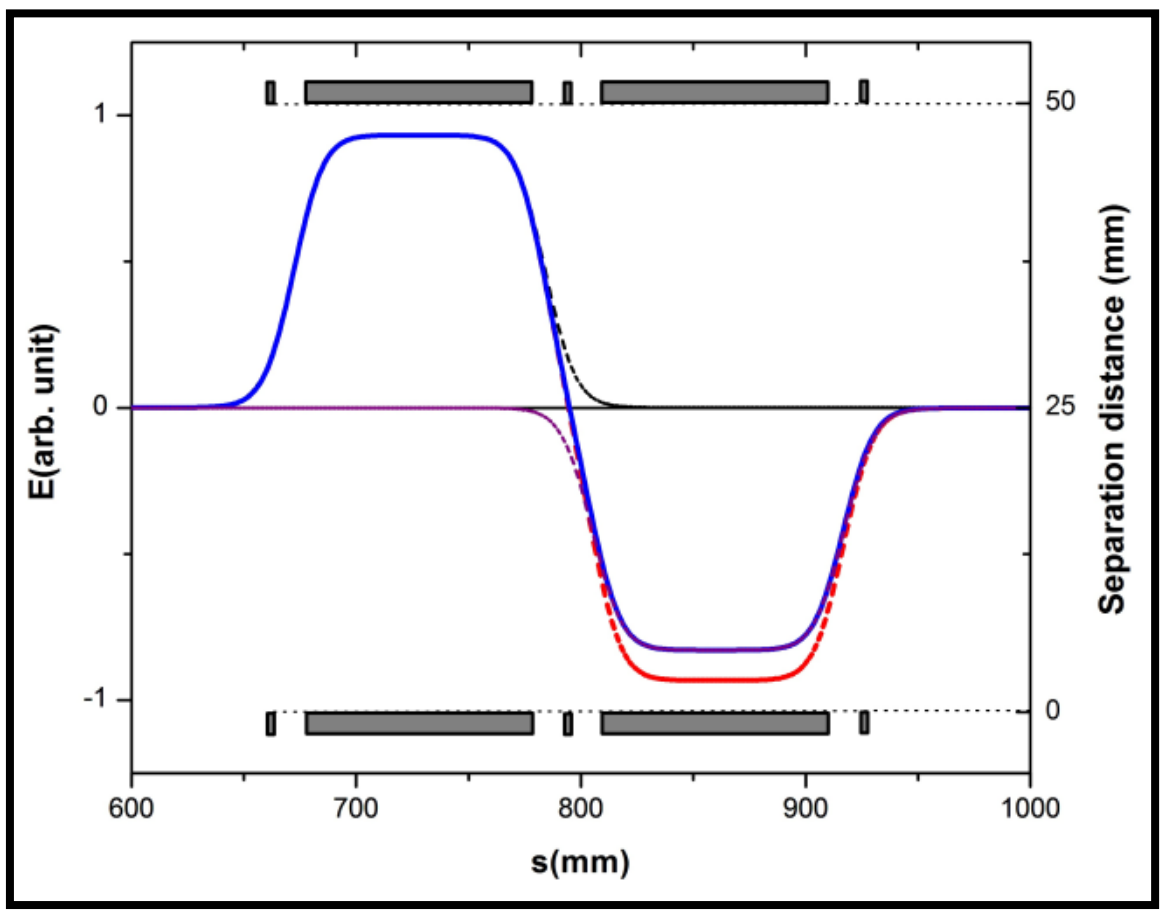

Figure 5: Field distributions in the quadrupole doublet at an off-axis position of x=5*mm* above the nominal orbit, as calculated by SIMION for the different voltages settings ($V_{QF}$, $V_{QD}$) in units of Volt. The blue curve shows one of the working points with a voltage setting of (405, 360), identical absolute voltages of (405, 405) are used for the red curve, a single focusing quadrupole (405, 0) is used for the dashed red curve and a single defocusing quadrupole (0, 360) is illustrated by the dashed blue curve.

Electrostatic quadrupoles generate aberrations of octupole order [36, 46], which can potentially disturb a traversing particle beam. For this reason, it is of high interest to calculate the aberrations induced by the ring`s quadrupoles and thus evaluate their significance with respect to beam divergence and energy spread. Several methods for calculating the aberrations of electrostatic lenses are described in literature [47, 48], with an excellent study on the subject by Sise, *et al* [35]. The latter carried out ray-tracing simulations using the SIMION and LENSYS programs, and investigated the spherical and chromatic aberration coefficients of multi-element electrostatic lens systems in great detail. A similar method has been applied in the present study. Chromatic and spherical aberrations have been calculated for the quadrupole, using the ray-tracing simulation program SIMION [49, 50].

As an example, the results shall be illustrated for the case where voltages of $V_{QF}$ = 405 V are used at a beam energy of 30 KeV. Assuming that particles originate from a point source at position *O*, which is the median point between two quadrupole doublets, the evolution of these aberrations as a function of the beam divergence angle *α*, defined at origin *O*, is shown in Fig. 7 and follows a parabola:

$$\Delta r = -MC_s\alpha^3 \qquad (10)$$

Here, $\Delta r$ is the radius of the disk of least confusion induced by spherical aberrations, *M* is the linear magnification of the lens, $\alpha$ is the half-angle of the beam divergence and $C_s$ is the spherical aberration coefficient, which depends on the ion optical system. A fit on the value of the radius of the aberration disk as function of $\alpha$ then allows us to define how these aberrations grow with the half-angle of beam divergence and how important they are, in particular at the maximum value of beam divergence. In the case studied here, the factor of the parabolic growth of spherical aberrations is about 11,100 mm/radian$^3$, yielding thus an aberration disc radius $\Delta r=200\mu m$ for $\alpha=1.5^o$, which is the maximum half-angle of the beam divergence assumed also for all other calculations such as the MAD-X lattice calculations.

Chromatic aberrations have been calculated in a similar way, again assuming a maximum beam divergence of $\alpha=1.5^o$ and their evolution as a function of relative energy spread of the beam $\Delta T/T$ is shown in Fig. 7. The energy spread $\Delta T$ goes up to 20 eV, which is a typical value for the energy spread from a standard Penning ion source like the one that will be used for initial operation of the ring. This figure shows a linear evolution of chromatic aberrations in agreement with:

$$\delta r = -MC_c\alpha\frac{\Delta T}{T} \qquad (11)$$

Here $\delta r$ is the radius of the disk of least confusion induced by chromatic aberrations. A fit to the values of the radius of the aberration disk as function of $\Delta T/T$, allows us to define how chromatic aberrations grow with a change in kinetic energy of the particles and how significant they can be. In the case discussed here, the linear growth factor of the chromatic aberrations is about *44.1 mm*, yielding an aberration disc radius of $\delta r = 30\ \mu m$ for $\Delta T = 20\ eV$. This does not necessary mean that spherical aberrations are dominant over chromatic ones, since the value of $\Delta T = 20\ eV$ only allows us to estimate the "natural" chromaticity and may be far lower than the change in energy induced by the electrostatic bends.

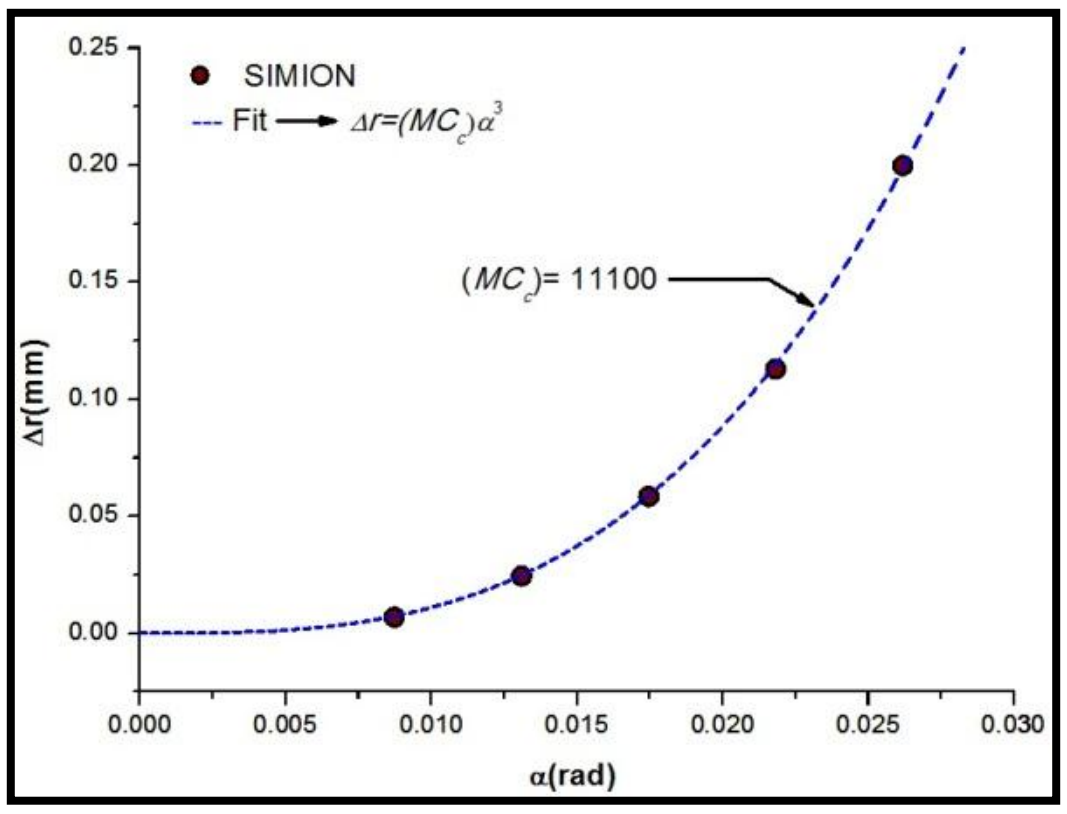

Figure 6: Evolution of spherical aberrations induced by the quadrupole. The radius of the aberration disc *Δr* is plotted as a function of the half angle α of the beam divergence at origin *O*. The aberrations disc is calculated in the range $0.5^o < \alpha < 1.5^o$, using the simulation program SIMION.

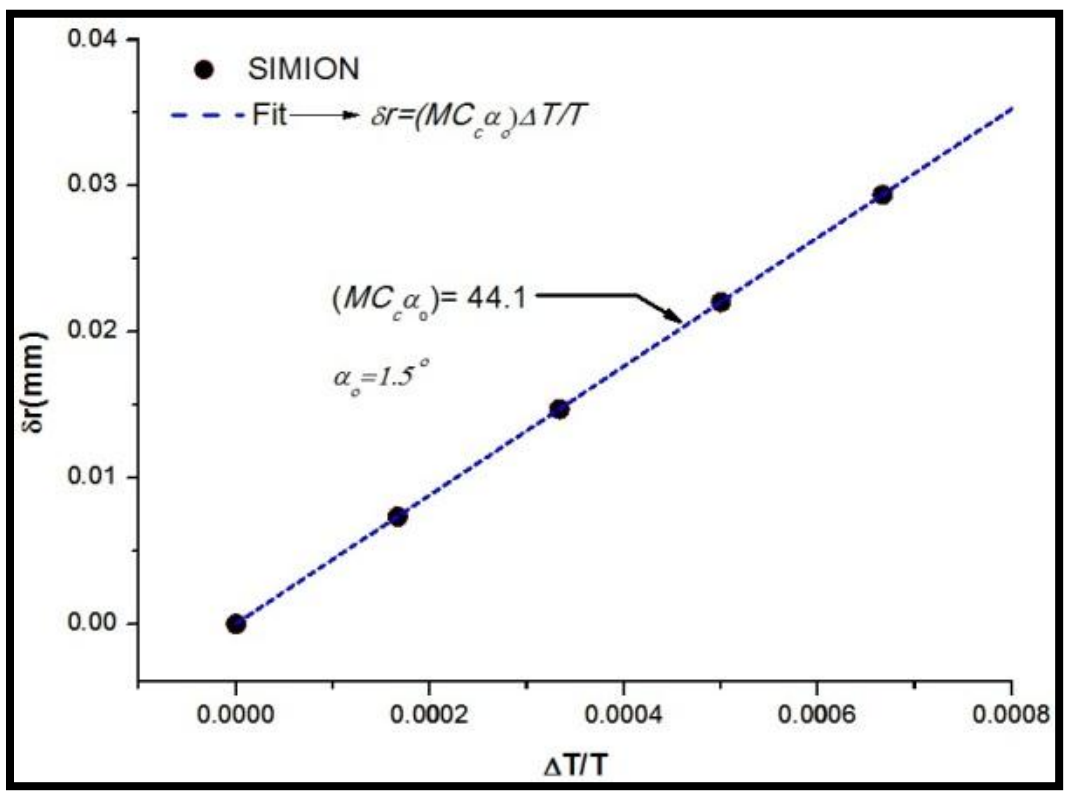

Figure 7: Evolution of chromatic aberrations, induced by the quadrupole. The radius of the aberration disc *δr* is plotted as a function of *ΔT/T*, with *ΔT* ranging from 0 to 20 eV. The plot is calculated using the SIMION program.

## 1. Storage Ring Layout

### 4.1 Split-bend Layout

The initial layout of the KACST storage ring consists of four super periods with double mirror symmetry, as illustrated in Fig. 8. Each period constitutes one quarter of the ring and this design was chosen to allow for a future upgrade of the ring to a double storage ring structure [31]. In this design, the bending is split into two identical bending corners, consisting of a 76° cylinder deflector (76° CD), equidistantly inserted between two 7° parallel-plate deflectors (7° PPDs). These form a 90° bending corner, connected on one side to a doublet of focusing and defocusing quadrupoles (QF, QD) and on the other side to a quadrupole singlet (QS). A vertical steerer (VS) is designed to be mounted on each side of the 76° CD for closed-orbit corrections in the vertical plane. The space between the different optical elements is kept field-free and is designed to accommodate beam diagnostics devices and experimental setups. From a beam dynamics point of view a lattice with only one central quadrupole singlet, equidistantly inserted between two 7° PPDs in the short straight section of the ring, would also work. However, it was decided to split this focusing action into two adjacent quadrupole singlets to be able to insert an experiment in the middle of the short straight section where the beam can be focused down to a small diameter [43]. A comprehensive beam dynamics study of this split-bending deflector layout goes beyond the scope of this paper and shall be the subject of further investigations that will include the effects from the non-linearity of the fields. Here, we limit the study to the single-bend design.

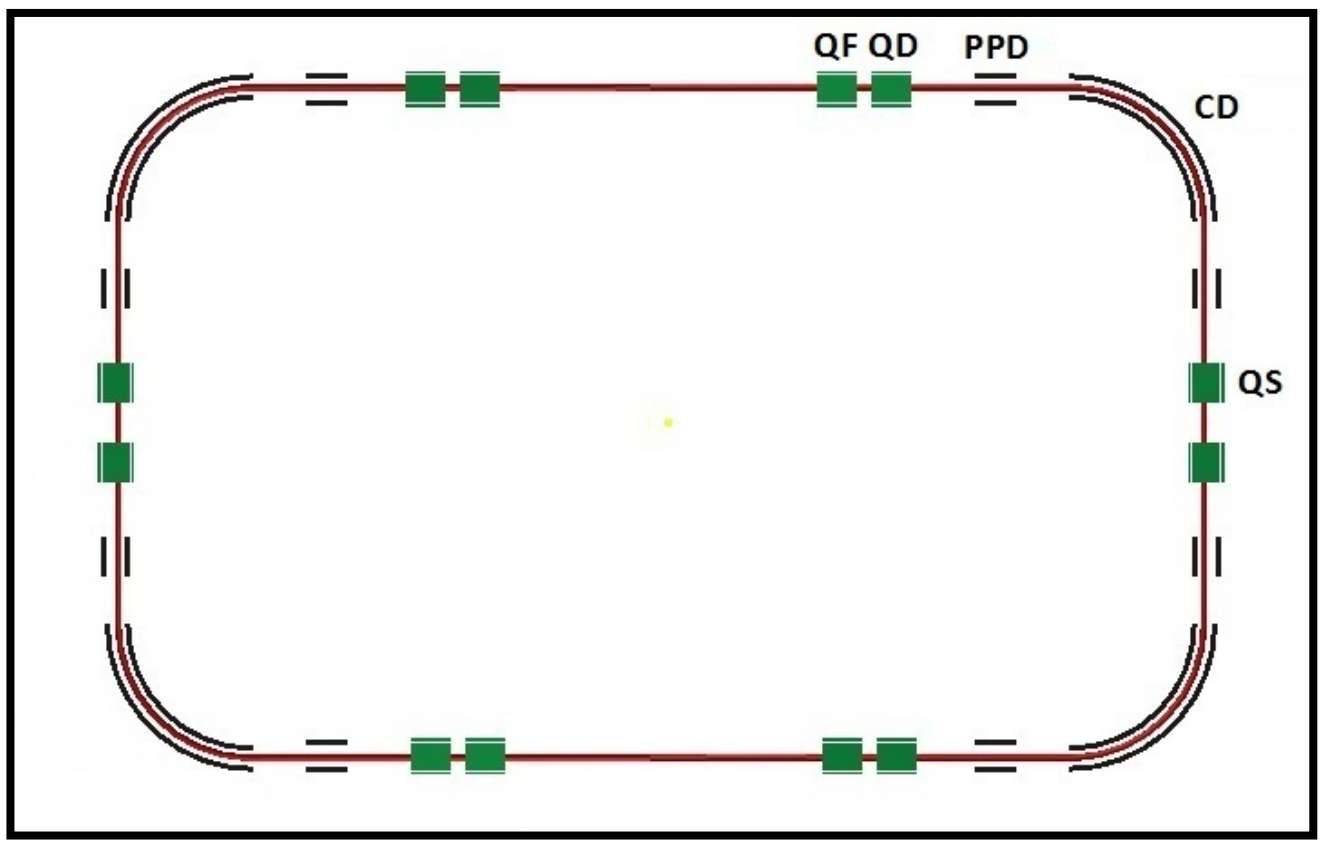

Figure 8: Lattice of the split-bend layout. QF: Focusing Quadrupole, QD: Defocusing Quadrupole, PPD: Parallel-Plate Deflector, CD: the Cylinder Deflector, (QS) Quadrupole Singlet.

### 4.2 Single-bend Racetrack Layout

For practical reasons, priority is given to a quick realization of a simple and early-run adaptation of the ring, followed by a later upgrade to the ultimate split-bend layout of the ring. In terms of design, this results in a simplification of the lattice presented in the previous section to a racetrack layout with a single 166° bend and two 7° Parallel-Plate Deflectors (7° PPDs) , see Fig. 9. This configuration avoids the need of any quadrupole singlet and significantly reduces the required number of ion optical elements. Thereby, it also shrinks the size of the ring, also simplifying certain beam dynamics studies and e.g. allowing the whole storage ring to be incorporated into a single instance of the SIMION simulation [43]. This configuration comes at the expense of slightly reducing the flexibility of the facility with regard to experimental opportunities. In the following, the description will be limited to this ring layout.

The configuration just described leads to a racetrack shaped ring with a circumference of roughly 9.2 m. Its detailed design is shown in Fig. 9. This single-bend racetrack design composes of two long straight sections that are connected by a U-turn bend that consists of a 166° CD, inserted in between two 7° PPDs. Closed-orbit corrections are realized by means of four vertical steerers (VS), designed to be located in the field-free region between the parallel-plate and cylindrical deflectors. Each straight section contains two quadrupole doublets (QF, QD), two field-free regions reserved for beam diagnostics, a central experimental section, 900 mm in length, for merged-beams experiments, and two shorter experimental sections 300mm long for crossed-beam experiments. The latter are adjacent to the 7° PPD thus enlarging the angular acceptance of the detector at the end of the long straight section and hence improving the collection of strongly divergent particles. The central section would even be large enough to integrate a small electron cooler similar to the one in the KEK-ring. The main design parameters of this layout are summarized in Table 1.

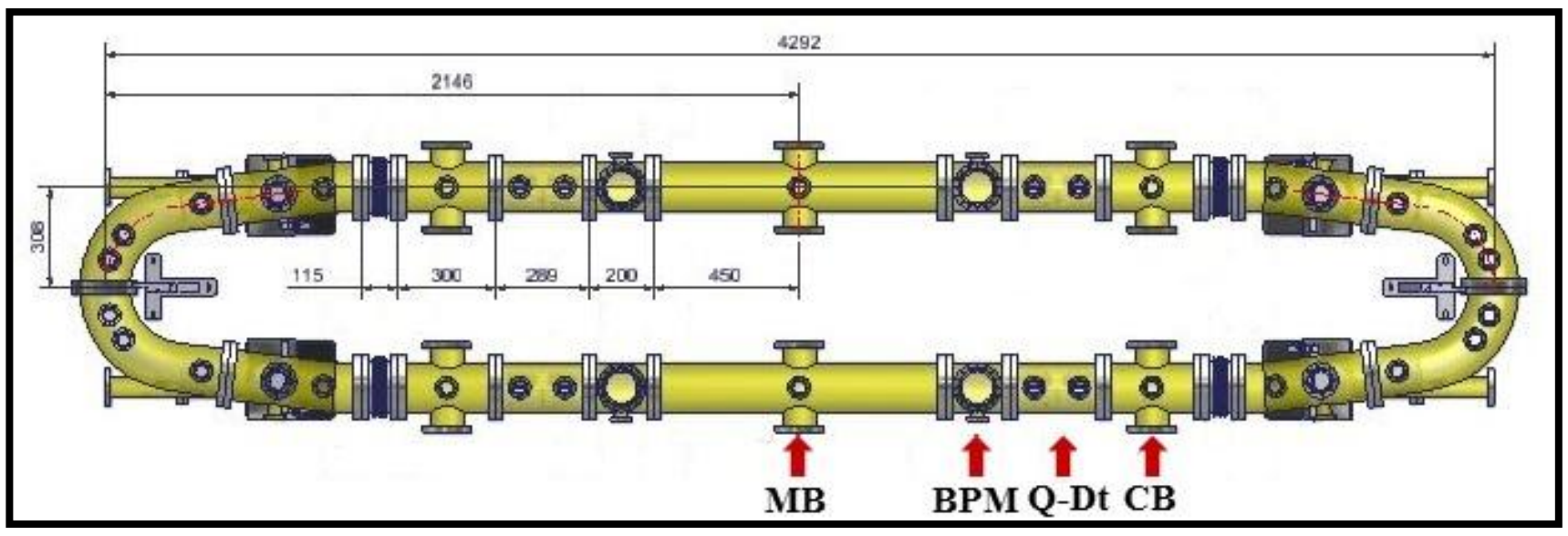

Figure 9.: 7°single-bend racetrack layout, as designed by Inventor 3 of AutoDesk.

In this configuration, the beam will be injected into the ring through one of the 7° Parallel-Plate Deflectors (PPDs) that would be switched off during injection and switched on right after single shot filling of the ring. The rise time of the field in the small angle deflector is of the order of 10% of the revolution time of the beam in the ring and hence should not have negative effects on the beam. Transverse modulation of the beam will be realized by four quadrupoles doublets (QF, QD) and parallel-plate vertical steerers will be used for closed-orbit corrections.

Table 1: Design parameters of the 7°single-bend racetrack layout.

| Parameters | Nominal value |
|---|---|
| Nominal beam energy [KeV·q] | 30 |
| Circumference [m] | 9.2 |
| Revolution time of protons[μs] | 3.8 |
| **Parallel-Plate Deflector** | |
| Nominal Deflection angle [°] | 7 |
| Plate area [mm×mm] | 100×100 |
| Electrode separation [mm] | 50 |
| Nominal voltage [kV] | ± 1.8 |
| **Cylinder Deflector** | |
| Nominal deflection angle [°] | 166 |
| Height [mm] | 100 |
| Central radius [mm] | 250 |
| Electrode separation [mm] | 30 |
| Electrodes-shield separation | 10mm |
| Nominal voltage [KV] | ± 3.6 |
| **Quadrupole Doublet** | |
| Inscribed radius [mm] | 25 |
| electrode length [mm] | 100 |
| Separation gap [mm] | 33 |
| Nominal voltage [kV] | ± 1 |

## 2. Beam Tracking - Stability

The SIMION simulation program allows charged-particles to be tracked through the real 3D field distribution, including non-linear effects from e.g. the fringe fields. In preliminary studies [31, 51] the different components of the lattice were individually modeled in separate geometry files and ions were then tracked through half of the ring lattice, using SIMION's IOB operation [50]. This approach leads to some inconsistencies in overlapping regions. Due to the limited extent of the individual geometries it also somehow idealized the field distributions and could not take fringe fields and other non-linear effects correctly into account. Consequently, it was not possible to track ions through the real field and 'store' them in the simulated ring, nor to study beam stability, lifetime and other important effects in detail. Here, the geometry was built on the basis of the ring's triple mirror symmetry [52]. This enables the size of the geometry file to be significantly reduced and the whole storage ring to be represented and incorporated in a single instance, with one point per millimeter resolution. The only effect that cannot be correctly implemented in this case is the action of the vertical steerers, as they require voltages of different polarity on either electrode. In addition, SIMION's in-built geometry feature for more complex geometries have been exploited to accurately model the cylindrical deflector including its longitudinal and lateral shields and to provide the possibility to optimize the effective lengths  of the CD. It should be pointed out that the single-bend racetrack design is small enough that it was even possible to include all vacuum chambers in the design and to thus track ions through the most realistic distribution of the fields. In this way, ions were successfully "stored" in the simulated ring for tens of thousands of turns, see Fig. 10. First simulation runs were performed to probe the ability of the ring to store low-energy beams, and a 30 KeV proton beam of diameters up 15 mm in the centre of the long straight section was stably stored in spite of fringe fields, i.e. without any adjustment of the geometrical length of the CD`s electrodes. Further simulation runs were then performed with an optimized lattice in which the cylindrical deflector setup including its shield was adjusted according to formula (9). In both cases the number of turns was very large and no difference between the two lattices in terms of beam stability was visible. However, limitations in ring acceptance and dynamic aperture were clearly observed in the case of the non-optimized lattice where fringe fields have not been adequately constrained. A more detailed analysis of these effects is the subject of further studies [53].

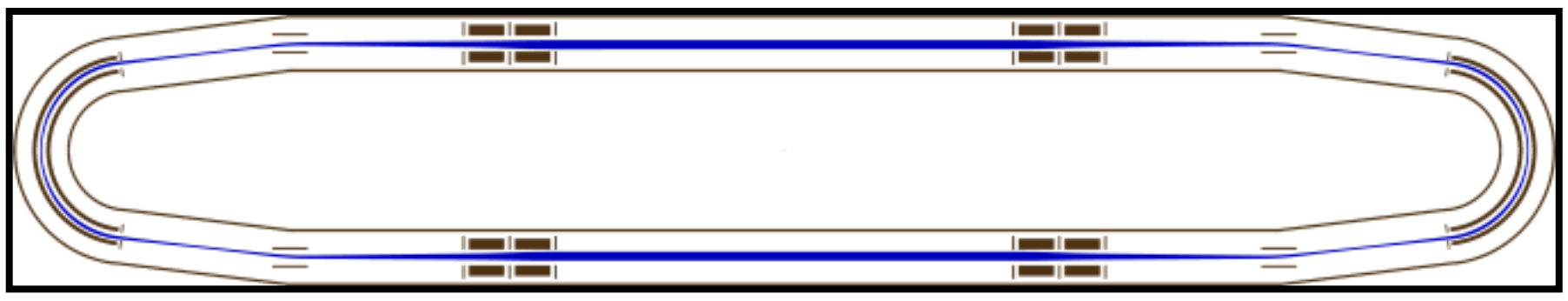

Figure 10: Cut-view of the simulated storage ring, in SIMION. In blue: a beam of 15 mm diameter is stored in the ring for thousands of turns.

In additional studies, this geometry allowed us modify and optimize the voltage settings on all electrodes, using the SIMION in-built fast adjust scaling routine. Tuning the quadrupoles voltages over a wide range enabled numerical studies into beam stability as a function of working points to be made. Such advanced studies also allow investigations to be performed into the ring's dynamic aperture, effects from different shield geometries or the impact of positioning and field errors.

By tracking ions through the single-bend racetrack lattice and changing $V_{QF}$ and $V_{QD}$, four stable regions were identified in our simulations around the working points A(405, 360), B(604, 716), C(585, 441), D (716, 760), which are in units of volts. Scanning over a range of quadrupole voltages around each working point allows us to probe and map the theoretical stable regions of the simulated storage ring. Such manual scans in a SIMION simulation, however, are very time consuming, and go beyond the scope of this paper. A detailed study will be reported separately [53].

## 3. Lattice Functions

In its original form, MAD-X does not support electrostatic deflectors and quadrupoles. The code allows, however, the R-matrix of an arbitrary beam optical element to be implemented and lattice studies carried out similar to those normally realized for magnetic rings. Also, MAD-X has been initially written for high energy, i.e. relativistic beams, and therefore does not includes fringe fields and non-linear effects that become much more important at low beam energies. These effects have been recently studied in detail [54]. The results obtained from MAD-X, COSY Infinity and a custom-written tracking routine relying on the field maps from the OPERA3D code were benchmarked against each other, to show the limitations of each method. Likewise, the here-presented simulation makes use of the custom-written tracking routine in order to determine the working points and stability regions in the single-bend racetrack design and to compare them to the results obtained from SIMION.

Preliminary lattice calculations have been done for the basic conceptual design [32-33]. However, the practices adopted for fringe fields limitation such as e.g. the addition of shields and others geometrical constraints and experimental requirements, imposed some changes to the optics configuration in such a conceptual design. Consequently, the lattice calculations must be revised. Here, we refined and reproduced the lattice function, using the MAD-X code for the ultimate single-bend racetrack design. Although only few quadrupoles are used in the single-bend racetrack design, the simulated ring can be set to many different operating modes, including elliptical and round beams, as illustrated in Fig. 11.

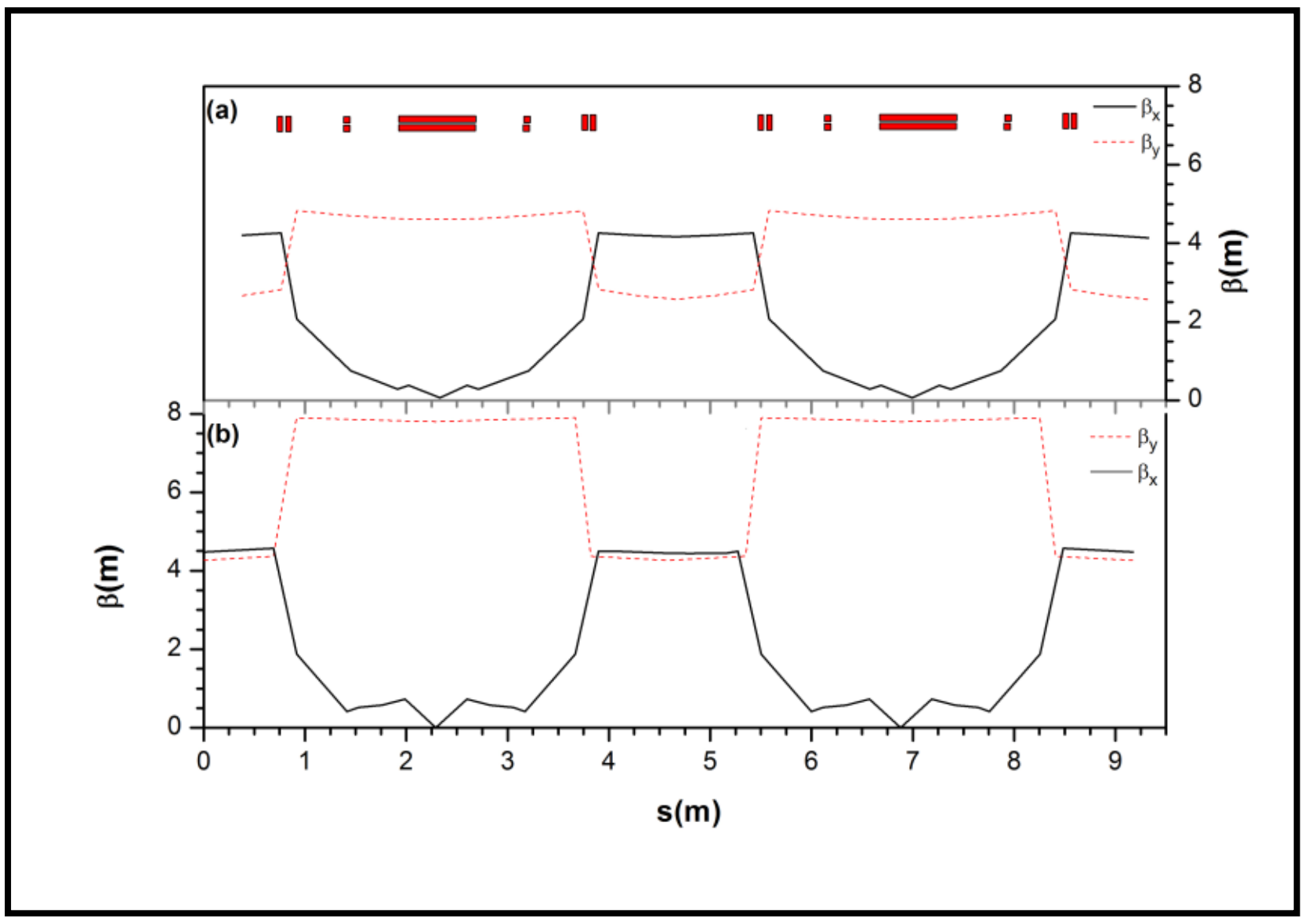

Figure 11: Two different operating modes of the simulated single-bend racetrack ring as calculated using MAD-X code. Top (a): Round beam in the centre of the straight section, optimized for merged/crossed beam experiments. Bottom (b): Elliptical beam in the centre of the straight section which keeps beam dimensions minimum all around the storage ring.

To numerically compute and delimit the theoretical stability regions of the single-bend racetrack design, a custom-written tracking routine relying on the field maps from OPERA3D [54] has been used. This routine models the ion trajectories through realistic field distributions and thus allows the stability of the beam to be probed. Fig. 12 shows the theoretical stability regions as a function of Focusing (F) and Defocusing (D) quadrupole-strength, $k_F$ and $k_D$. The results from the tracking routine agree with those carried out using the SIMION program, which track ions in the most realistic fields. In fact, one can see that each of the afore-mentioned working points expressed in term of quadrupole strengths ($k_F$ , $k_D$) as A(2.16, 1.92), B(1.71, 1.28) , C(3.30, 1.72), D (3.65, 2.32), in units of $m^2$, fits into a corresponding stability region. Analysis of detailed effects clearly showed that the results from the tracking studies provided much more detailed results than those that could be obtained by using MAD-X. In some cases, limitations of using an R-Matrix based approach when studying beam stability and dynamic aperture were clearly observed and priority should be given to studies based on the real three-dimensional field distribution.

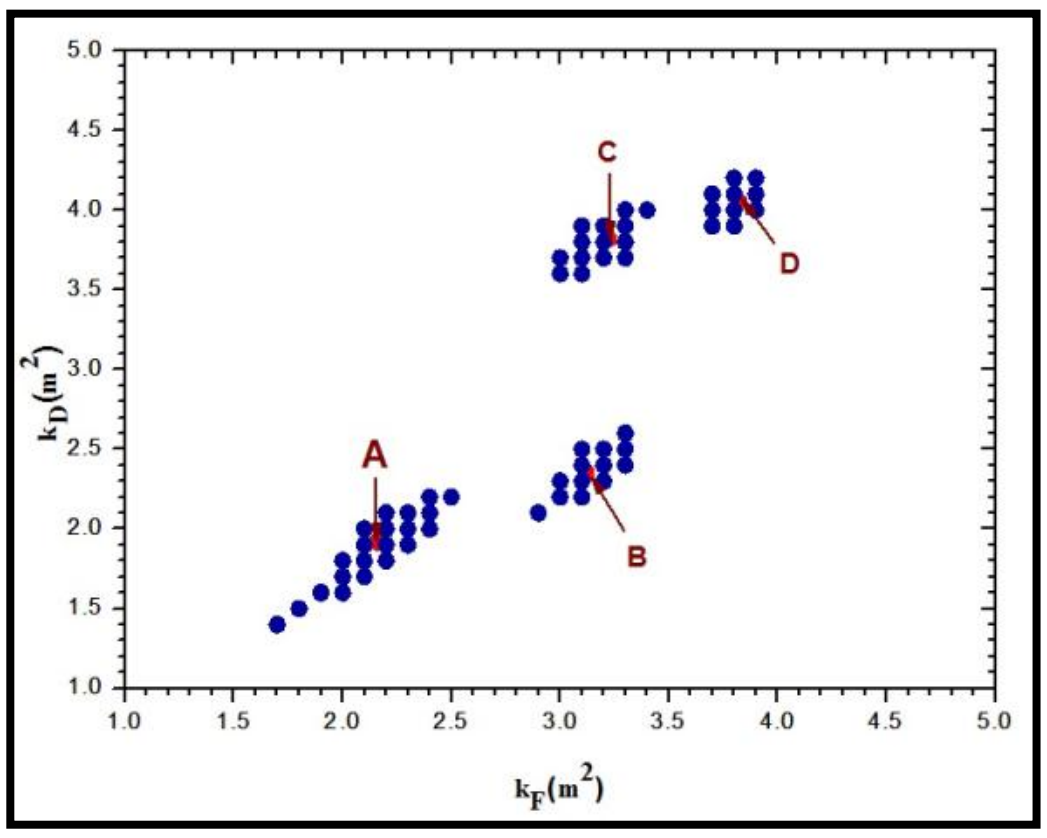

Figure 12: The four working points A-D and corresponding stability regions, as calculated using the custom-written tracking routine for single-bend racetrack design. The plot is represented in a ($k_F$, $k_D$) diagram where $k_F$/$k_D$ are the absolute value of the quadrupole strength.

Table 2.: Lattice parameters of the 7$^o$ single-bend layout for an elliptic beam.

| Parameters | Value |
|---|---|
| Emittance [π mm mrad] | 15 |
| Acceptance [π mm mrad] | 25 |
| **Lattice Functions** | |
| $\beta_{x,\ initial}$ [m] | 4.2 |
| $\beta_{y,\ initial}$ [m] | 2.6 |
| $\beta_{x,\ min}$ [m] | 0.1 |
| $\beta_{y,\ max}$ [m] | 4.8 |
| Nominal voltage [kV] | ±3.6 |
| **Betatron Tunes** | |
| $Q_x$ | *.74* |
| $Q_y$ | *.2* |
| **Chromaticity** | |
| $\xi_x$ | -3.1 |
| $\xi_y$ | -2.2 |
| **Quadrupole Strengths** | |
| $k_{QF}$ [$m^2$] | 2.2 |
| $k_{QD}$ [$m^2$] | -1.8 |

## 4. Conclusion

The design of a 7$^o$-electrostatic storage ring that can store ions with energies of up to 30keV.q and that is being in development at KACST, was presented in this paper. Although this experimental facility shall consist of a double ring configuration in its final phase, allowing for a very broad and interdisciplinary research program, it was decided to start with a simpler, but quicker to realize 'start version'. This small storage ring will already offer many exciting research opportunities for local researchers as well as for the international research community. In this paper details on the 7$^o$-design of the storage ring were presented with an emphasis on an accurate simulation of the effects from non-linear fields and the results from simulation studies with SIMION. The present design study has complemented and finalized the design of the 7$^o$ single-bend racetrack-shaped electrostatic storage ring at KACST.

**Acknowledgments**

M. O. A. El Ghazaly who initiated and leads this project for building the electrostatic storage ring at KACST (Pro-ESR), gratefully thanks S P Møller for comments related to SIMION implementations and advice in designing the electrostatic storage ring. Valuable discussions with P. Defrance are also gratefully acknowledged. Thanks go to Henrik Juul at the University of Aarhus for help in the mechanical design, and to the NCMP technicians at KACST for their efforts related to this work. This project is funded by KACST under the grant no. 162-28/MOA_El Ghazaly. M.O.A El Ghazaly (project PI) gratefully thanks KACST for this grant.